\documentclass[12pt]{article}

\usepackage{epsfig}			

\newcommand{\eras}[0]{{\cal T}}
\newcommand{\numeras}[0]{{|\eras|}}
\newcommand{\bolddot}[0]{\mbox{$\cdot$}}
\newcommand{\xdot}[0]{x_{\bolddot}}

\newcommand{\zn}[0]{{Z_{1,N}}}
\newcommand{\zeras}[0]{{Z_{1,\numeras}}}
\newcommand{\zex}[0]{{Z_{ext}}}
\newcommand{\zin}[0]{{Z_{int}}}
\newcommand{\bbar}[0]{\ol{B}}

\newcommand{\beq}{\begin{equation}}  
\newcommand{\eeq}{\end{equation}}
\newcommand{\beqa}{\begin{eqnarray*}}  
\newcommand{\eeqa}{\end{eqnarray*}}

\newcommand{\ul}[1]{\underline{#1}}
\newcommand{\ol}[1]{\overline{#1}}
\newcommand{\enote}[1]{\cite{#1}}    
\newcommand{\eqlabel}[1]{\renewcommand{\theequation}{#1}}

\begin{document}
\title{How to Compile \\ A Quantum Bayesian Net}

\author{Robert R. Tucci\\
        P.O. Box 226\\ 
        Bedford,  MA   01730\\
        tucci@ar-tiste.com}

\date{ \today} 

\maketitle

\vskip2cm
\section*{Abstract}
We show how to express the information contained in a Quantum Bayesian (QB) net 
as a product of unitary matrices. If each of these unitary matrices is 
expressed as a sequence of elementary operations (operations such as controlled-nots and
qubit rotations), then the result is a sequence of operations that
can be used to run a quantum computer. QB nets have been run entirely on 
a classical computer, but one expects them to run faster on a quantum computer.
\newpage
\subsection*{Introduction}
\mbox{}\indent

Quantum Bayesian (QB) Nets\enote{Tucci95}-\enote{QFog} are a method of modelling quantum
systems graphically in terms of network diagrams.  In this
paper, we show how to 
express the information contained in a QB net
as a product of unitary operators.
In another paper\enote{Tucci98a},
we have presented a method for
reducing any unitary operator to a sequence of elementary operations (SEO).
Such SEOs can be used to 
manipulate an array of quantum bits, a quantum computer.\enote{DiV}-\enote{ECorrection}
Thus, combining the results of this paper and those
of Ref.\enote{Tucci98a}, one can reduce a
QB net into a SEO that
can be used to run a quantum computer.
Of course, QB nets 
can and have been run entirely on a classical computer\enote{QFog}. However, because
of the higher speeds promised by quantum parallelism, one expects
QB nets to run much faster on a quantum computer.

The process of reducing a QB net to a SEO is analogous to the 
process of ``compiling source code" for classical computers.
With classical computers, one writes a computer program in a high 
level language (like Fortran, C or C++). A compiler then expresses
this as a SEO for manipulating bits. In the case
of quantum computers, a QB net may be thought of as a program in a
high level language. In the future, programs called ``quantum compilers"
will be widely available. These will run (initially) on classical computers.
They  will take whatever QB net we give it, and 
re-express it as a SEO
that  will then be used to control an array of quantum bits.
These quantum compilers will take the QB net entered by the user
and automatically add to it quantum error correction code\enote{ECorrection}
and various speed optimizations.
	
QB nets are to quantum physics what Classical Bayesian (CB) nets\enote{CBnets}  
are to classical physics. CB nets have been used very successfully in the field of 
artificial intelligence (AI). In fact, even Microsoft's Office Suite 
contains CB net code\enote{Micro$oft}. Thus, we hope and expect that some day QB nets,
running on quantum computers, will be used for AI applications. In fact,
we believe that quantum computers are ideally suited for such applications. First, 
because AI tasks often require tremendous power, and quantum computers seem to 
promise this. Second, because quantum computers are plagued by quantum noise,
which makes their coherence times short. There are palliatives to this, such 
as quantum error correction\enote{ECorrection}. But such palliatives come at 
a price: a large increase in the number of steps. 
The current literature often mentions 
factoring a large number into primes\enote{factoring}
as a future use of quantum computers.
However, due to noise, 
quantum computers may ultimately prove to be impractical for doing 
long precise calculations such as this. On the other hand, short
coherence times appear to be a less serious problem for 
the types of calculations
involved in AI. The human brain has coherence times too short to factor
a 100 digit number into primes, and yet long enough to
conceive the frescoes in the Sistine Chapel. We do not mean to imply that
the human brain is a quantum computer.  An airplane is not a bird, but 
it makes a good flyer. Perhaps a quantum computer, although not a
human brain, can make a good thinker.

\subsection*{Review}
\mbox{}\indent

We begin by presenting a brief review of QB nets. For more information,
see Ref.\enote{Tucci95}-\enote{QFog}.

In what follows, we use the following notation. We define 
$Z_{a, b} = \{ a, a+1, \ldots, b\}$ for any integers $a$ and $b$.
$\delta(x,y)$ equals one if $x=y$ and zero otherwise.
For any finite set $S$, $|S|$ denotes the number of elements
in $S$.

We call a {\it graph} (or a diagram ) a collection of nodes with arrows connecting some
pairs of these nodes. The arrows of the graph must satisfy certain
constraints. We call a {\it labelled graph}
a graph whose nodes are
labelled.  A
{\it QB net}  consists of  two parts: a  labelled graph with each node
labelled by a random variable,  and a collection of node matrices, one 
matrix for each node. These two parts must satisfy certain constraints.

An {\it internal arrow} is an arrow that
has  a starting (source) node and a different ending (destination) one.
We will use only internal arrows.
We  define two
types of nodes:  an {\it internal node} is a node that has
one or more internal arrows leaving it, and an {\it external node} is a node that has no
internal arrows leaving it. It is also common to
use the terms {\it root node} or {\it prior probability node} for a node
which has no incoming arrows (if any arrows touch it, they are outgoing ones). 

We restrict our attention to {\it acyclic} graphs; that is, graphs that do not contain cycles.
(A {\it cycle} is a closed path of arrows with the arrows all pointing in the same sense.)

 We assign a
 random variable to each node of the QB net. (Henceforth, we will underline random
variables.  For example, we might  write $P(\ul{x}=x)$ for the probability that the
random variable $\ul{x}$ assumes the particular value
$x$.) Suppose the random
variables assigned to the $N$ nodes are $\ul{x}_1,\ul{x}_2,\cdots,\ul{x}_N$. For each $j\in \zn$, 
the random variable $\ul{x}_j$ will be assumed to take on
values  within a finite set $\Sigma_j$ called
{\it the set of possible states of} $\ul{x}_j$. 

If $S=\{k_1,k_2,\cdots, k_{|S|}\}\subset \zn$,
and $k_1< k_2 < \cdots < k_{|S|}$,
define $(x_{\bolddot})_S=(x_{k_1},x_{k_2},\cdots,x_{k_{|S|}})$ and 
$(\ul{x}_{\bolddot})_S=(\ul{x}_{k_1},\ul{x}_{k_2},\cdots,\ul{x}_{k_{|S|}})$.
Sometimes, we also abbreviate $(x_{\bolddot})_{\zn}$ 
(i.e., the vector that includes all the possible $x_j$ components) by just
$x_{\bolddot}$, and $(\ul{x}_{\bolddot})_{\zn}$ by just $\ul{x}_{\bolddot}\;$.

Let $\zex$ be the set of all $j\in \zn$ such that $\ul{x}_j$ is
an external node,  and  let $\zin$ be the set of all $j\in \zn$ such
that 
 $\ul{x}_j$  is an internal node. Clearly, $\zex$ and $\zin$ are disjoint
and their union is $\zn$. 

Each possible value $x_{\bolddot}$ of 
$\ul{x}_{\bolddot}$ defines a different {\it net story}.
 For any net story $\xdot$,
we call $(\xdot)_\zin$ the {\it internal state of the story} and 
$(\xdot)_\zex$ its {\it external state}.
 
For each net story, we may assign an amplitude to each node.
Define $S_j$ to be the set of all $k$ such that 
an arrow labelled $x_k$ (i.e., an arrow whose source node is $\ul{x}_k$) enters node 
$\ul{x}_j$. 
We
assign a complex number $A_j[x_j|(x_{\bolddot})_{S_j}]$ to node $\ul{x}_j$.
 We call
$A_j[x_j|(x_{\bolddot})_{S_j}]$  the {\it amplitude of node} $\ul{x}_j$
{\it within net story} $x_{\bolddot}$. 

The {\it amplitude of net story} $\xdot$,
call it $A(\xdot)$,  is  defined to 
be the product of all the node amplitudes 
$A_j[x_j|(x_{\bolddot})_{S_j}]$ for  $j\in \zn$. Thus,

\beq
A(\xdot)=\prod_{j\in \zn}
A_j[x_j|(x_{\bolddot})_{S_j}]
\;.
\eqlabel{1}\eeq 

The function $A_j$ with values $A_j[x_j|(x_{\bolddot})_{S_j}]$ determines a matrix that we 
call the  {\it node matrix of node} $\ul{x}_j$.
 $x_j$ is the matrix's  {\it row index} and   $(x_{\bolddot})_{S_j}$ is
its {\it column index}.

\subsection*{Method}
\mbox{}\indent

		\begin{center}
			\epsfig{file=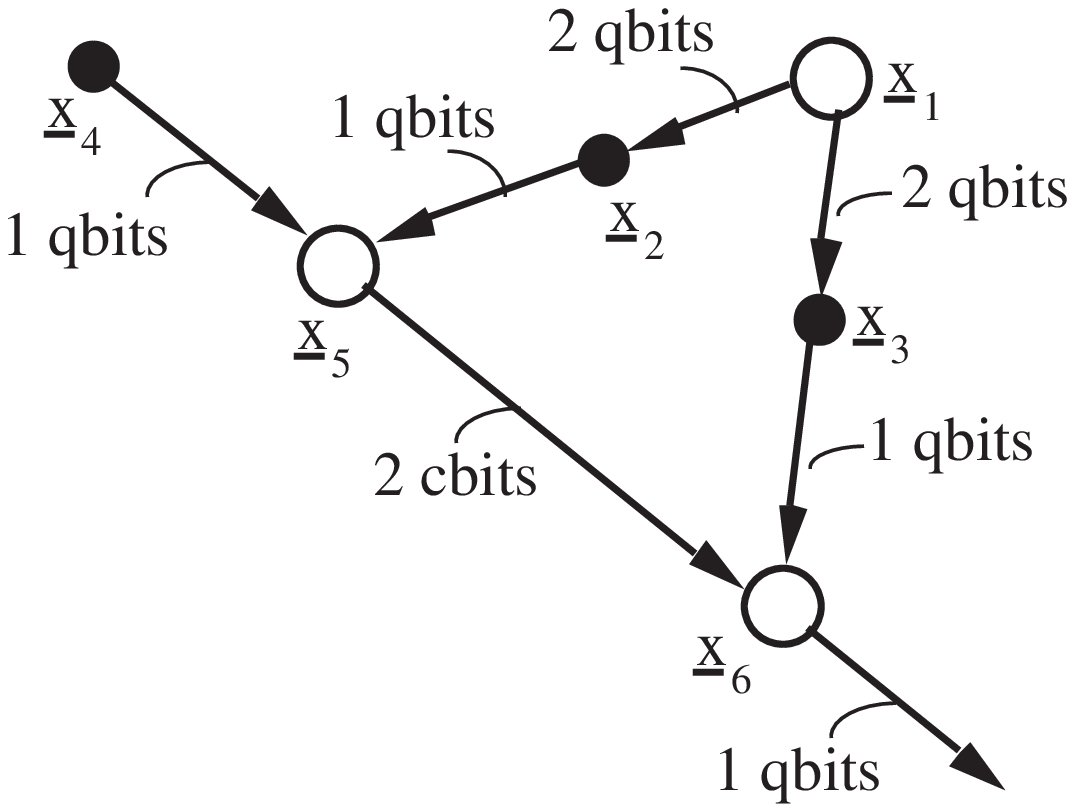}
			
			{Fig.1 QB net for Teleportation.
			This figure also shows
			the number of quantum or classical bits carried by each arrow. }
		\end{center}


One can translate a QB net into a SEO by performing 
the following 3 steps: (1) Find eras, (2) Insert delta functions, (3) Find unitary 
extensions of era matrices. Next we will discuss these steps in detail. We will
illustrate our discussion by using Teleportation \enote{Teli} as an example.
Figure 1 shows a QB net for Teleporation. This net is discussed in Ref.\enote{QFog}.
Reference \enote{Bra} gives
a SEO, expressed graphically 
as a qubit circuit, for Teleportation.  It appears that the 
author of  Ref.\enote{Bra}
obtained his circuit mostly by hand,
based on  information very similar to that contained in a QB net.
This paper gives a general method
whereby such  circuits can be obtained from a QB net in a completely mechanical way
by means of a classical computer.

\subsection*{\bf Step 1: Find eras}
\mbox{}\indent

		\begin{center}
			\epsfig{file=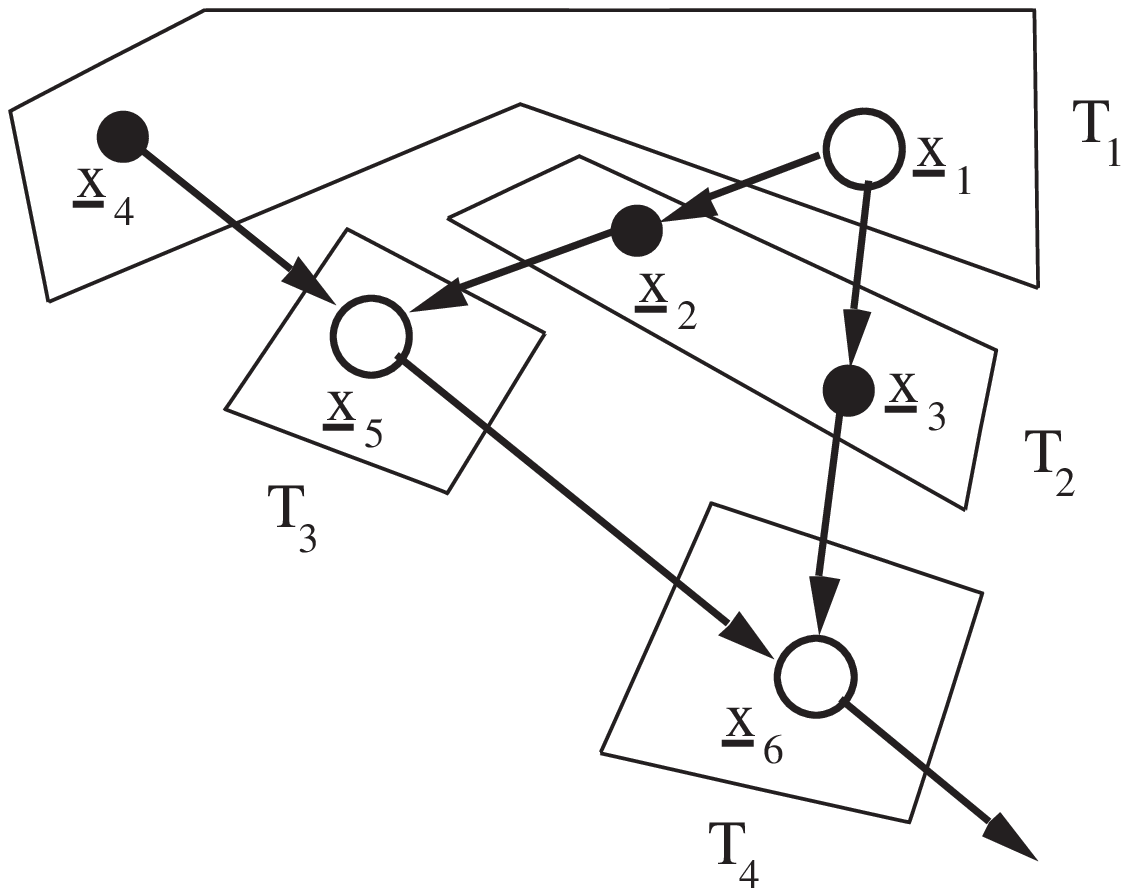}
			
			{Fig.2 Root node eras for Teleportation net.}
		\end{center}

	
The {\it root node eras} of a graph are defined as follows.
Call the original graph Graph(1). The first era $T_1$ is defined as
the set of all root nodes of Graph(1).
Call Graph(2) the graph obtained by erasing 
from Graph(1) all the $T_1$ nodes and any arrows connected to these nodes. 
Then $T_{2}$ is defined as the set of all root nodes 
of Graph(2). One can continue this process until one defines an era $T_\numeras$
such that Graph($\numeras + 1$) is empty. (One can show that 
if Graph(1) is acyclic, then one always arrives at a Graph($\numeras + 1$) that is empty.) 
 For example, Fig.2 shows the
root node eras for the Teleportation net Fig.1.
Let $\eras$ represent the set of eras: $\eras = \{T_1, T_2, \cdots, T_\numeras\}$.
Note that $T_a\subset \zn $ for all $a\in \zeras$ and 
the union of all $T_a$ equals $\zn$. In mathematical parlance,
the collection of eras is a partition of $\zn$.

		\begin{center}
			\epsfig{file=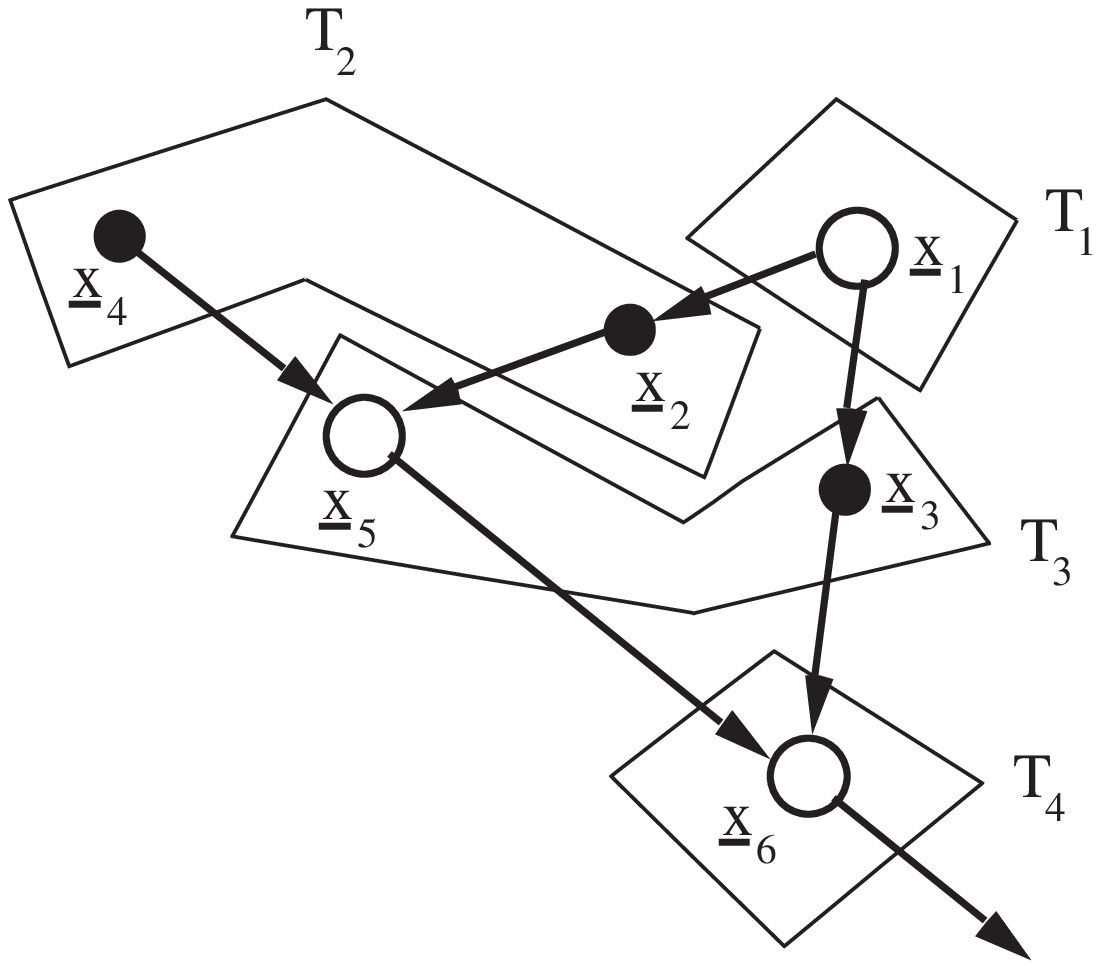}
			
			{Fig.3 External node eras for Teleportation net.}
		\end{center}


Rather than defining eras by (1) removing successive layers of root nodes, one can
also define them by (2) removing successive layers of external nodes.
We call this second type of era, the {\it external node eras} of the graph.
For example, Fig.3 shows the external node eras of the Teleportation net Fig.1. 

This process whereby one classifies the nodes of an acyclic graph into 
eras is a well know technique referred to as a  chronological or
topological sort in the computer literature\enote{Flamig}.

Henceforth, for the sake of
definiteness, we will speak only of root node eras. The case of 
external node eras can be treated similarly.

Suppose that $a\in \zeras$. The arrows exiting the $a$'th era are labelled by 
$(\xdot)_{T_a}$. Those entering it are labelled by $(\xdot)_{\Gamma_a}$, where $\Gamma_a$
is defined by
$\Gamma_a = \bigcup_{j \in T_a} S_j$.
Note that the $a$'th era node is only entered by arrows 
from nodes that belong to previous (not subsequent) eras 
so $\Gamma_a \subset T_{a-1}\cup \ldots \cup T_2 \cup T_1$. 
The amplitude $B_a$ of the $a$'th era is defined as

\beq
B_a[ (\xdot)_{T_a} | (\xdot)_{\Gamma_a} ] =
\prod_{j\in T_a}
A_j[ x_j | (\xdot)_{S_j}]
\;.
\eqlabel{2}\eeq
The amplitude $A(\xdot)$ of story $\xdot$ is given by

\beq
A(\xdot) = \prod_{a = 1}^{\numeras} B_a
\;.
\eqlabel{3}\eeq

For example, for Teleportation we get from Fig.2

\beq
B_1(x_1, x_4) = A_1(x_1) A_4(x_4)
\;,
\eqlabel{4a}\eeq

\beq
B_2(x_2, x_3 | x_1) = A_2(x_2 | x_1) A_3(x_3 | x_1)
\;,
\eqlabel{4b}\eeq

\beq
B_3(x_5 | x_2, x_4) = A_5(x_5 | x_2, x_4 )
\;,
\eqlabel{4c}\eeq

\beq
B_4(x_6 | x_3, x_5) = A_6(x_6 | x_3, x_5 )
\;,
\eqlabel{4d}\eeq

and
\beq
A(\xdot) = B_4 B_3 B_2 B_1
\;.
\eqlabel{5}\eeq

\subsection*{Step 2: Insert delta functions}
\mbox{}\indent	

The Feynman Integral $FI$ for a QB net is  defined by

\beq
FI[ (\xdot)_\zex ] = \sum_{ (\xdot)_\zin } A(\xdot)
\;.
\eqlabel{6}\eeq
Note that we are summing over all stories $\xdot$ that have 
$(\xdot)_\zex$ as their external state.
We want to express the right side of Eq.(6) as a 
product of matrices.

Consider how to do this for Teleportation. In that case one has

\beq
FI(x_6) = \sum_{x_1, x_2, \ldots x_5} B_4 B_3 B_2 B_1
\;,
\eqlabel{7}\eeq
where the $B_a$ are given by Eqs(4).
The right side of Eq.(7) is not ready to be expressed as a product of
 matrices because the column indices of $B_{a+1}$ and the row 
indices of $B_a$ are not the same for all $a\in Z_{1, \numeras -1}$.
Furthermore, the variable $x_3$ occurs in $B_4$ and $B_2$
but not in $B_3$. Likewise, the variable $x_4$ occurs
in $B_3$ and $B_1$ but not in $B_2$.
Suppose we define
$\bbar_a$ for $a\in \zeras$ by

\beq
\bbar_1(x_1^1, x_4^1) = B_1(x_1^1, x_4^1)
\;,
\eqlabel{8a}\eeq

\beq
\bbar_2(x_2^2, x_3^2, x_4^2 | x_1^1, x_4^1) = B_2(x_2^2, x_3^2 | x_1^1)
\delta(x_4^2, x_4^1)
\;,
\eqlabel{8b}\eeq

\beq
\bbar_3(x_3^3, x_5^3 | x_2^2, x_3^2, x_4^2) = B_3(x_5^3 | x_2^2, x_4^2)
\delta(x_3^3, x_3^2)
\;,
\eqlabel{8c}\eeq

\beq
\bbar_4(x_6 | x_3^3, x_5^3) = B_4(x_6 | x_3^3, x_5^3)
\;.
\eqlabel{8d}\eeq
Then 

\beq
FI(x_6) = \sum_{interm} \bbar_4 \bbar_3 \bbar_2 \bbar_1
\;,
\eqlabel{9}\eeq
where we sum over all intermediate indices; i.e., all $x_j^a$ except $x_6$.
Contrary to Eq.(7), the right side of Eq.(9) can be 
expressed immediately as a product of matrices since now 
$B_{a+1}$ column indices and $B_a$ row indices are the same.
The purpose of 
inserting a delta function of $x_3$ into $B_3$ is to
allow the
system to ``remember" the value of $x_3$ between non-consecutive eras $T_4$ and 
$T_2$.
Inserting a delta function of $x_4$ into $B_2$ serves a similar purpose.

		\begin{center}
			\epsfig{file=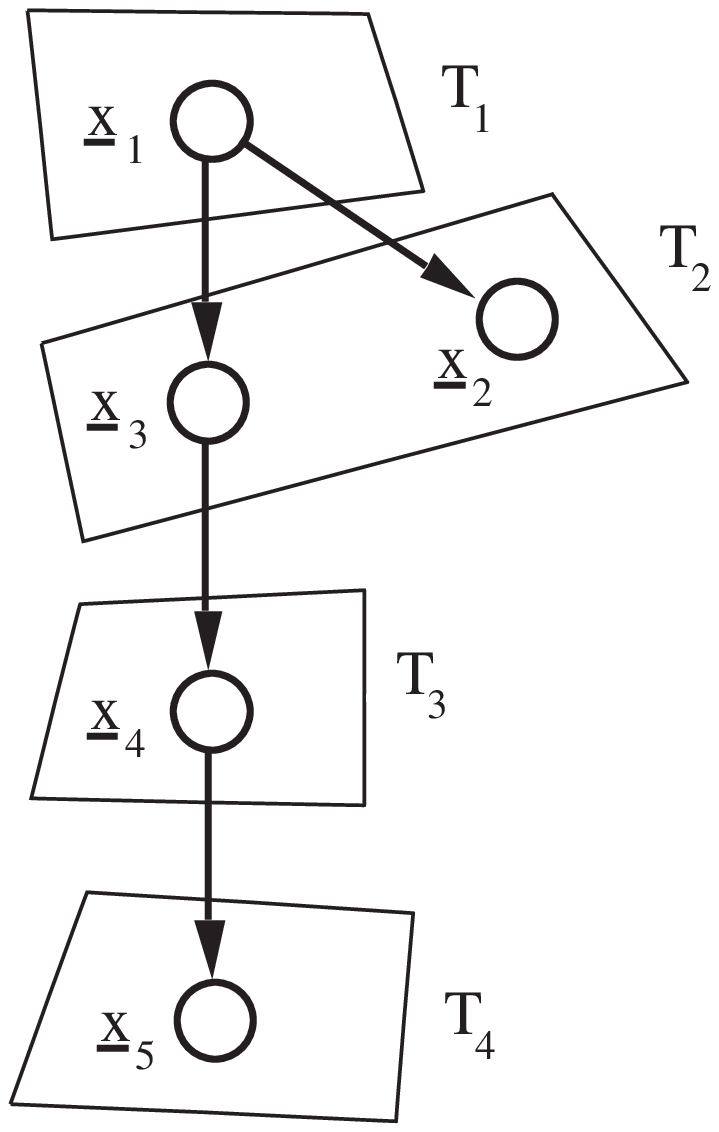}
			
			{Fig.4 Example of a QB net in which 
			an external node is not in the final era.}
		\end{center}


In the Teleporation net of Fig.1, the last era contains all the external
nodes. However, for some QB nets like the one in Fig.4, this is not the case.
For the net of Fig.4,

\beq
B_1(x_1) = A_1(x_1)
\;,
\eqlabel{10a}\eeq

\beq
B_2(x_2, x_3 | x_1) = A_3(x_3 | x_1) A_2(x_2 | x_1) 
\;,
\eqlabel{10b}\eeq

\beq
B_3(x_4 | x_3) = A_4(x_4 | x_3 )
\;,
\eqlabel{10c}\eeq

\beq
B_4(x_5 | x_4) = A_5(x_5 | x_4 )
\;.
\eqlabel{10d}\eeq
Even though node $\ul{x}_2$ is external,
the variable $x_2$ does not appear as a row index in $B_4$.
Suppose we set

\beq
\bbar_1(x_1^1) = B_1(x_1^1)
\;,
\eqlabel{11a}\eeq

\beq
\bbar_2(x_2^2, x_3^2 | x_1^1) = B_2(x_2^2, x_3^2 | x_1^1)
\;,
\eqlabel{11b}\eeq

\beq
\bbar_3(x_2^3, x_4^3 | x_2^2, x_3^2,) = B_3(x_4^3 | x_3^2)
\delta(x_2^3, x_2^2)
\;,
\eqlabel{11c}\eeq

\beq
\bbar_4(x_2, x_5 | x_2^3, x_4^3) = B_4(x_5 | x_4)
\delta(x_2, x_2^3)
\;.
\eqlabel{11d}\eeq
Then 

\beq
FI(x_2, x_5) = \sum_{interm} \bbar_4 \bbar_3 \bbar_2 \bbar_1
\;,
\eqlabel{12}\eeq
where we sum over all intermediate indices; i.e., all $x_j^a$ except $x_2$ and $x_5$. 
Contrary to $B_4$, 
the rows of $\bbar_4$ are labelled by the indices of both external nodes $\ul{x}_2$ and $\ul{x}_5$.

This technique of inserting delta functions
can be generalized as follows to deal with arbitrary QB nets. For
$j\in \zn$, let $a_{min}(j)$ be the smallest $a\in \zeras$ such that 
$x_j$ appears in $B_a$. Hence, $a_{min}(j)$ is the first era in which 
$x_j$ appears.  If $\ul{x}_j$ is an internal node,
 let $a_{max}(j)$ be the largest $a$ such that 
$x_j$ appears in $B_a$ (i.e., the last era in which $x_j$ appears). If 
$\ul{x}_j$ is an external node, let $a_{max}(j) = \numeras + 1$.
For $a\in \zeras$, let

\beq
\Delta_a = \{ j\in \zn | a_{min}(j) < a < a_{max}(j) \}
\;,
\eqlabel{13}\eeq

\beq
\bbar_a = B_a[ (\xdot^a)_{T_a} | (\xdot^{a-1})_{\Gamma_a} ]
\prod_{j \in \Delta_a} \delta (x^a_j, x^{a-1}_j )
\;.
\eqlabel{14}\eeq
In Eq.(14), $x^\numeras_j$ should be identified with $x_j$ and 
$x^0_j$ with no variable at all. 
Equation(6) for $FI$ can be written in terms of the $\bbar_a$ functions:

\beq
FI[ (\xdot)_\zex ] = \sum_{interm} \bbar_\numeras \ldots \bbar_2 \bbar_1
\;,
\eqlabel{15}\eeq
where the sum is over all intermediate indices (i.e., all $x^a_j$
for which $a\neq \numeras$).
For all $a$, define 
matrix
$M_a$ so that the $x,y$
entry of $M_a$ is $\bbar_a(x|y)$. 
Define $M$ to be a column vector whose components are the values of $FI$ for each external state.
Then Eq.(15) can be expressed as:

\beq
M = M_\numeras \ldots M_2 M_1
\;.
\eqlabel{16}\eeq
The rows of the column vector $M$ are labelled by the 
possible values of $(\xdot)_\zex$. The rows of the 
column vector $M_1$ are labelled by the possible values of
$(\xdot)_{T_1}$, where $T_1$ is the set of root nodes.

\subsection*{\bf Step 3: Find unitary 
extensions of era matrices}
\mbox{}\indent	
So far, we have succeeded in expressing $FI$ as a product
of matrices $M_a$, but these matrices are not necessarily 
unitary. 
In this step, we will show how to extend each $M_a$ matrix
(by adding rows and columns) into a unitary matrix $U_a$. The techniques of 
Ref.\enote{Tucci98a} will then be applicable to each matrix $U_a$. 

 By combining adjacent $M_a$'s, one can 
produce a new, smaller set of matrices $M_a$. Suppose the  
union of two consecutive eras is also defined to be an era. 
Then combining adjacent $M_a$'s is equivalent
to combining two consecutive eras to produce a new, smaller set of eras.
We define a {\it breakpoint} as any position 
$a\in Z_{1, \numeras -1}$ between two adjacent matrices 
$M_{a+1}$ and $M_a$.
Combining two adjacent $M_a$'s eliminates a breakpoint. Breakpoints
are only necessary at positions 
where internal measurements are made. For example,
in Teleportation experiments, one measures
node $\ul{x}_3$, which is in era $T_3$.
Hence, a breakpoint between $M_4$ and $M_3$ is necessary.
If that is the only internal measurement to be made, all
other breakpoints can be dispensed with.
Then we will have $M = M'_2 M'_1$ where
$M'_2 = M_4$, $M'_1 = M_3 M_2 M_1$.
If no internal measurements are made, then
we can combine all matrices $M_a$ into a single one,
and eliminate all breakpoints.

We will henceforth assume that 
for all $a\in \zeras$, the columns of $M_a$ are orthonormal.
If for some $a_0\in \zeras$, $M_{a_0}$ does not satify this condition,
it may be possible to ``repair" $M_{a_0}$ so that it does.
First:
If a row $\beta$
of $M_{a_0 - 1 }$ is zero, then eliminate the column $\beta$
of $M_{a_0}$, and the row $\beta$ of $M_{a_0 -1}$.
Next: If a row $\beta$ of the column vector 
$M_{a_0-1}\ldots M_2 M_1$ is zero, then flag the column $\beta$ of $M_{a_0}$. 
The flagged columns of $M_{a_0}$
can be changed without affecting the value of $M$.
If the non-flagged columns of $M_{a_0}$ are orthonormal,
and the number of columns in $M_{a_0}$ does not exceed 
the number of rows, then 
the Gram Schmidt method, to be discussed later, can be used
to replace the flagged columns by new columns such that all the 
columns of the new matrix 
$M_{a_0}$ are orthonormal. 
If it is not possible to repair 
$M_{a_0}$ in any of the above ways (or 
in some other way that might become clear
once we program this), one can always 
remove the breakpoint between
$M_{a_0 + 1}$ and $M_{a_0}$.

		\begin{center}
			\epsfig{file=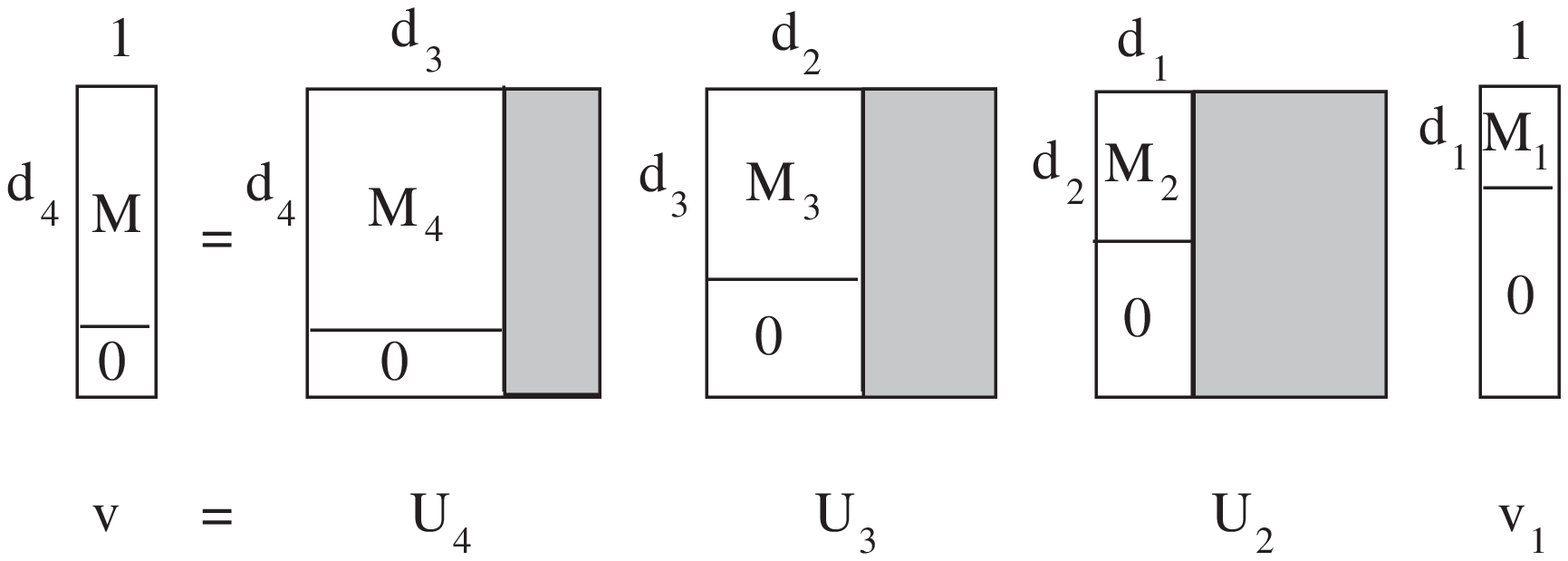}
			
			{Fig.5  Dimensions of matrices $M_a$ and of 
			their unitary extensions $U_a$.}
		\end{center}


Consider Fig.5. We will call $d_{a}$  the number
of rows of matrix $M_a$ and $d_{a-1}$ its number of columns. We define $D$ and $N_S$ by

\beq
D= \max\{ d_a | 1\leq a \leq \numeras \} 
\,
\eqlabel{17}\eeq

\beq
N_S = \min \{
2^i | i\in Z_{1, \infty}, D \leq 2^i \}
\;.
\eqlabel{18}\eeq
Let $\ol{d}_a = N_S - d_a$ for all $a$.For each 
$a\neq 1$, we define $U_a$ to be the matrix that
one obtains by extending $M_a$ as follows. We 
append an $\ol{d}_a \times d_{a-1}$ block of zeros
beneath $M_a$ and an $N_S \times \ol{d}_{a-1}$
block of gray entries to the right of $M_a$. 
By gray entries we mean entries whose value is yet
to be specified. 
When $a=1$, $M_1$ be can extended in two ways.
One can append a column of $\ol{d}_1$
zeros beneath it and call the resulting $N_S$ dimensional column vector $v_1$.
Alterntatively, one can append a column of  
$\ol{d}_1$
zeros beneath $M_1$ and 
an $N_S \times (N_S - 1)$
block of gray entries to the right of $M_1$, 
and
call the resulting $N_S \times N_S$
matrix $U_1$. In this second case, one must also insert
$e_1$ to the right of $U_1$.
By $e_1$ we mean the $N_S$ dimensional column vector
whose first entry equals one and all others equal zero.
Which extension of $M_1$ is used, whether the one that
requires $e_1$ or the one that doesn't, 
should be left as a choice of the user.
Henceforth, for the sake of
definiteness, we will 
will assume  that the user has chosen the
extension without the $e_1$. The other case can be treated similarly.
Equation(16) then becomes

\beq
v = U_\numeras \ldots U_3 U_2 v_1
\;,
\eqlabel{19}\eeq
where $v$ is just the column vector
$M$ with $\ol{d}_\numeras$ zeros attached to the end.
Note that 

\beq
U_a \ldots U_2 v_1 =
\left [
\begin{array}{c}
M_a \ldots M_2 M_1 \\
0
\end{array}
\right ]
\;,
\eqlabel{20}\eeq
for all $a\in \zeras$, where the zero
indicates a column of  $\ol{d}_a$
zeros.

To determine suitable values for the gray entries of the $U_a$ matrices, one can 
use the Gram-Schmidt (G.S.) method \enote{Noble}. This method
takes as input an ordered set $S = (v_1, v_2,\ldots, v_N)$
of vectors, not necessarily independent ones.
It yields as output another ordered set 
of vectors
$S' = (u_1, u_2,\ldots, u_N)$,
such that $S'$ spans the same vector space as $S$.
Some vectors in $S'$ may be zero. Those vectors
of $S'$ which aren't zero will be orthonormal.
For $r\in \zn$, if the first $r$ vectors
of $S$ are already orthonormal, then the
first $r$ vectors of $S'$ will be the 
same as the first $r$ vectors of $S$. Let
$e_j$ for $j\in Z_{1, N_S}$ be the $j$'th
standard unit vector (i.e., the vector whose  $j$'th
entry is one and all other entries are zero).
For 
each $a\in \zeras$,
to determine the gray entries of $U_a$  one can use the G.S. method
on the set $S$ consisting of the non-gray columns of
$U_a$ together with the vectors $e_1, e_2, \ldots e_{N_S}$.

\newpage
\section*{FIGURE CAPTIONS:}
\begin{description}

\item{\sc Fig.1} QB net for Teleportation.
This figure also shows
the number of quantum or classical bits carried by each arrow. 

\item{\sc Fig.2} Root node eras for Teleportation net.

\item{\sc Fig.3} External node eras for Teleportation net.

\item{\sc Fig.4} Example of a QB net in which 
an external node is not in the final era.

\item{\sc Fig.5} Dimensions of matrices $M_a$ and of 
their unitary extensions $U_a$.

\end{description}
\end{document}